\documentstyle[12pt]{article}

\catcode`\@=11
\long\def\@makefntext#1{ 
\protect\noindent \hbox to 3.2pt {\hskip-.9pt
$^{{\ninerm\@thefnmark}}$\hfil}#1\hfill} 

\def\thefootnote{\fnsymbol{footnote}}
 \def\@makefnmark{\hbox to 0pt{$^{\@thefnmark}$\hss}}  

\def\ps@myheadings{\let\@mkboth\@gobbletwo
\def\@oddhead{\hbox{} 
\rightmark\hfil\ninerm\thepage}
\def\@oddfoot{}\def\@evenhead{\ninerm\thepage\hfil 
\leftmark\hbox{}}\def\@evenfoot{}
\def\sectionmark##1{}\def\subsectionmark##1{}}

\begin{document}

\newcommand{\symbolfootnote}{\renewcommand{\thefootnote}
	{\fnsymbol{footnote}}}
\renewcommand{\thefootnote}{\fnsymbol{footnote}}
\newcommand{\alphfootnote}
	{\setcounter{footnote}{0}
	 \renewcommand{\thefootnote}{\sevenrm\alph{footnote}}}

\newcounter{sectionc}\newcounter{subsectionc}\newcounter{subsubsectionc}
\renewcommand{\section}[1] {\vspace{0.6cm}\addtocounter{sectionc}{1}
\setcounter{subsectionc}{0}\setcounter{subsubsectionc}{0}\noindent
	{\bf\thesectionc. #1}\par\vspace{0.4cm}}
\renewcommand{\subsection}[1] {\vspace{0.6cm}\addtocounter{subsectionc}{1}
	\setcounter{subsubsectionc}{0}\noindent
	{\it\thesectionc.\thesubsectionc. #1}\par\vspace{0.4cm}}
\renewcommand{\subsubsection}[1]
{\vspace{0.6cm}\addtocounter{subsubsectionc}{1}
	\noindent {\rm\thesectionc.\thesubsectionc.\thesubsubsectionc.
	#1}\par\vspace{0.4cm}}
\newcommand{\nonumsection}[1] {\vspace{0.6cm}\noindent{\bf #1}
	\par\vspace{0.4cm}}

\newcounter{appendixc}
\newcounter{subappendixc}[appendixc]
\newcounter{subsubappendixc}[subappendixc]
\renewcommand{\thesubappendixc}{\Alph{appendixc}.\arabic{subappendixc}}
\renewcommand{\thesubsubappendixc}
	{\Alph{appendixc}.\arabic{subappendixc}.\arabic{subsubappendixc}}

\renewcommand{\appendix}[1] {\vspace{0.6cm}
        \refstepcounter{appendixc}
        \setcounter{figure}{0}
        \setcounter{table}{0}
        \setcounter{equation}{0}
        \renewcommand{\thefigure}{\Alph{appendixc}.\arabic{figure}}
        \renewcommand{\thetable}{\Alph{appendixc}.\arabic{table}}
        \renewcommand{\theappendixc}{\Alph{appendixc}}
        \renewcommand{\theequation}{\Alph{appendixc}.\arabic{equation}}
        \noindent{\bf Appendix \theappendixc #1}\par\vspace{0.4cm}}
\newcommand{\subappendix}[1] {\vspace{0.6cm}
        \refstepcounter{subappendixc}
        \noindent{\bf Appendix \thesubappendixc. #1}\par\vspace{0.4cm}}
\newcommand{\subsubappendix}[1] {\vspace{0.6cm}
        \refstepcounter{subsubappendixc}
        \noindent{\it Appendix \thesubsubappendixc. #1}
	\par\vspace{0.4cm}}

\def\abstracts#1{{
	\centering{\begin{minipage}{30pc}\tenrm\baselineskip=12pt\noindent
	\centerline{\tenrm ABSTRACT}\vspace{0.3cm}
	\parindent=0pt #1
	\end{minipage} }\par}}

\newcommand{\bibit}{\it}
\newcommand{\bibbf}{\bf}
\renewenvironment{thebibliography}[1]
	{\begin{list}{\arabic{enumi}.}
	{\usecounter{enumi}\setlength{\parsep}{0pt}
\setlength{\leftmargin 1.25cm}{\rightmargin 0pt}
	 \setlength{\itemsep}{0pt} \settowidth
	{\labelwidth}{#1.}\sloppy}}{\end{list}}

\topsep=0in\parsep=0in\itemsep=0in
\parindent=1.5pc

\newcounter{itemlistc}
\newcounter{romanlistc}
\newcounter{alphlistc}
\newcounter{arabiclistc}
\newenvironment{itemlist}
    	{\setcounter{itemlistc}{0}
	 \begin{list}{$\bullet$}
	{\usecounter{itemlistc}
	 \setlength{\parsep}{0pt}
	 \setlength{\itemsep}{0pt}}}{\end{list}}

\newenvironment{romanlist}
	{\setcounter{romanlistc}{0}
	 \begin{list}{$($\roman{romanlistc}$)$}
	{\usecounter{romanlistc}
	 \setlength{\parsep}{0pt}
	 \setlength{\itemsep}{0pt}}}{\end{list}}

\newenvironment{alphlist}
	{\setcounter{alphlistc}{0}
	 \begin{list}{$($\alph{alphlistc}$)$}
	{\usecounter{alphlistc}
	 \setlength{\parsep}{0pt}
	 \setlength{\itemsep}{0pt}}}{\end{list}}

\newenvironment{arabiclist}
	{\setcounter{arabiclistc}{0}
	 \begin{list}{\arabic{arabiclistc}}
	{\usecounter{arabiclistc}
	 \setlength{\parsep}{0pt}
	 \setlength{\itemsep}{0pt}}}{\end{list}}

\newcommand{\fcaption}[1]{
        \refstepcounter{figure}
        \setbox\@tempboxa = \hbox{\tenrm Fig.~\thefigure. #1}
        \ifdim \wd\@tempboxa > 6in
           {\begin{center}
        \parbox{6in}{\tenrm\baselineskip=12pt Fig.~\thefigure. #1 }
            \end{center}}
        \else
             {\begin{center}
             {\tenrm Fig.~\thefigure. #1}
              \end{center}}
        \fi}

\newcommand{\tcaption}[1]{
        \refstepcounter{table}
        \setbox\@tempboxa = \hbox{\tenrm Table~\thetable. #1}
        \ifdim \wd\@tempboxa > 6in
           {\begin{center}
        \parbox{6in}{\tenrm\baselineskip=12pt Table~\thetable. #1 }
            \end{center}}
        \else
             {\begin{center}
             {\tenrm Table~\thetable. #1}
              \end{center}}
        \fi}

\def\@citex[#1]#2{\if@filesw\immediate\write\@auxout
	{\string\citation{#2}}\fi
\def\@citea{}\@cite{\@for\@citeb:=#2\do
	{\@citea\def\@citea{,}\@ifundefined
	{b@\@citeb}{{\bf ?}\@warning
	{Citation `\@citeb' on page \thepage \space undefined}}
	{\csname b@\@citeb\endcsname}}}{#1}}
\newif\if@cghi
\def\cite{\@cghitrue\@ifnextchar [{\@tempswatrue
	\@citex}{\@tempswafalse\@citex[]}}
\def\@cite#1#2{{$\null^{#1}$\if@tempswa\typeout
	{IJCGA warning: optional citation argument
	ignored: `#2'} \fi}}
\newcommand{\citeup}{\cite}

\def\@citelowx[#1]#2{\if@filesw\immediate\write\@auxout
	{\string\citation{#2}}\fi
\def\@citelowa{}\@citelow{\@for\@citeb:=#2\do
	{\@citelowa\def\@citelowa{,}\@ifundefined
	{b@\@citeb}{{\bf ?}\@warning
	{Citation `\@citeb' on page \thepage \space undefined}}
	{\csname b@\@citeb\endcsname}}}{#1}}
\newif\if@cghi
\def\citelow{\@cghifalse\@ifnextchar [{\@tempswatrue
	\@citelowx}{\@tempswafalse\@citelowx[]}}
\def\@citelow#1#2{{$\null{[#1]}$\if@tempswa\typeout
	{IJCGA warning: optional citation argument
	ignored: `#2'} \fi}}
\def\fnm#1{$^{\mbox{\scriptsize #1}}$}
\def\fnt#1#2{\footnotetext{\kern-.3em
	{$^{\mbox{\sevenrm #1}}$}{#2}}}

\font\twelvebf=cmbx10 scaled\magstep 1
\font\twelverm=cmr10 scaled\magstep 1
\font\twelveit=cmti10 scaled\magstep 1
\font\elevenbfit=cmbxti10 scaled\magstephalf
\font\elevenbf=cmbx10 scaled\magstephalf
\font\elevenrm=cmr10 scaled\magstephalf
\font\elevenit=cmti10 scaled\magstephalf
\font\bfit=cmbxti10
\font\tenbf=cmbx10
\font\tenrm=cmr10
\font\tenit=cmti10
\font\ninebf=cmbx9
\font\ninerm=cmr9
\font\nineit=cmti9
\font\eightbf=cmbx8
\font\eightrm=cmr8
\font\eightit=cmti8
%
%
%
\def\JSP#1#2#3{{\sl J. Stat. Phys.} {\bf #1} (#2) #3}
\def\PRL#1#2#3{{\sl Phys. Rev. Lett.} {\bf#1} (#2) #3}
\def\PR#1#2#3{{\sl Phys. Rev.} {\bf#1} (#2) #3}
\def\EPL#1#2#3{{\sl Europhys. Lett.} {\bf#1} (#2) #3}
\def\NPB#1#2#3{{\sl Nucl. Phys.} {\bf B#1} (#2) #3}
\def\NPBFS#1#2#3#4{{\sl Nucl. Phys.} {\bf B#2} [FS#1] (#3) #4}
\def\CMP#1#2#3{{\sl Comm. Math. Phys.} {\bf #1} (#2) #3}
\def\CPAM#1#2#3{{\sl Comm. Pure Appl. Math.} {\bf #1} (#2) #3}
\def\PRD#1#2#3{{\sl Phys. Rev.} {\bf D#1} (#2) #3}
\def\PRB#1#2#3{{\sl Phys. Rev.} {\bf B#1} (#2) #3}
\def\PLB#1#2#3{{\sl Phys. Lett.} {\bf #1B} (#2) #3}
\def\PLA#1#2#3{{\sl Phys. Lett.} {\bf #1A} (#2) #3}
\def\JMP#1#2#3{{\sl J. Math. Phys.} {\bf #1} (#2) #3}
\def\JMM#1#2#3{{\sl J. Math. Mech.} {\bf #1} (#2) #3}
\def\PTP#1#2#3{{\sl Prog. Theor. Phys.} {\bf #1} (#2) #3}
\def\SPTP#1#2#3{{\sl Suppl. Prog. Theor. Phys.} {\bf #1} (#2) #3}
\def\AoP#1#2#3{{\sl Ann. of Phys.} {\bf #1} (#2) #3}
\def\APNY#1#2#3{{\sl Ann. Phys. (N.Y.)} {\bf #1} (#2) #3}
\def\PNAS#1#2#3{{\sl Proc. Natl. Acad. Sci. USA} {\bf #1} (#2) #3}
\def\RMP#1#2#3{{\sl Rev. Mod. Phys.} {\bf #1} (#2) #3}
\def\PRep#1#2#3{{\sl Phys. Reports} {\bf #1} (#2) #3}
\def\AoM#1#2#3{{\sl Ann. of Math.} {\bf #1} (#2) #3}
\def\UMN#1#2#3{{\sl Usp. Mat. Nauk} {\bf #1} (#2) #3}
\def\RMS#1#2#3{{\sl Russian Math Surveys} {\bf #1} (#2) #3}
\def\FAP#1#2#3{{\sl Funkt. Anal. Prilozheniya} {\bf #1} (#2) #3}
\def\FAaIP#1#2#3{{\sl Func. Anal. and Its Appl.} {\bf #1} (#2) #3}
\def\BSMF#1#2#3{{\sl Bull. Soc. Mat. France} {\bf #1} (#2) #3}
\def\BAMS#1#2#3{{\sl Bull. Am. Math. Soc.} {\bf #1} (#2) #3}
\def\TAMS#1#2#3{{\sl Trans. Am. Math. Soc.} {\bf #1} (#2) #3}
\def\AIHP#1#2#3{{\sl Ann. Inst. Henri Poincar\'e} {\bf #1} (#2) #3}
\def\ANYAS#1#2#3{{\sl Ann. New York Acad. Sci.} {\bf #1} (#2) #3}
\def\AIF#1#2#3#4{{\sl Ann. Inst. Fourier} {\bf #1,#2} (#3) #4}
\def\PAMS#1#2#3{{\sl Proc. Am. Math. Soc.} {\bf #1} (#2) #3}
\def\PRS#1#2#3{{\sl Proc. Roy. Soc.} {\bf #1} (#2) #3}
\def\CMJ#1#2#3{{\sl Czechosl. Math. J.} {\bf #1} (#2) #3}
\def\CompM#1#2#3{{\sl Compositio Math.} {\bf #1} (#2) #3}
\def\Compt#1#2#3{{\sl Compt. Rend. Acad. Sci. Paris} {\bf #1} (#2) #3}
\def\Invm#1#2#3{{\sl Invent. math.} {\bf #1} (#2) #3}
\def\LMP#1#2#3{{\sl Letters in Math. Phys.} {\bf #1} (#2) #3}
\def\IJMPA#1#2#3{{\sl Int. J. Mod. Phys.} {\bf A#1} (#2) #3}
\def\IJMPB#1#2#3{{\sl Int. J. Mod. Phys.} {\bf B#1} (#2) #3}
\def\AdM#1#2#3{{\sl Advances in Math.} {\bf #1} (#2) #3}
\def\AdP#1#2#3{{\sl Advances in Phys.} {\bf #1} (#2) #3}
\def\RMaP#1#2#3{{\sl Reports on Math. Phys.} {\bf #1} (#2) #3}
\def\IJM#1#2#3{{\sl Ill. J. Math.} {\bf #1} (#2) #3}
\def\APP#1#2#3{{\sl Acta Phys. Polon.} {\bf #1} (#2) #3}
\def\TMP#1#2#3{{\sl Theor. Mat. Phys.} {\bf #1} (#2) #3}
\def\JPA#1#2#3{{\sl J. Physics} {\bf A#1} (#2) #3}
\def\JPC#1#2#3{{\sl J. Physics} {\bf C#1} (#2) #3}
\def\JPCM#1#2#3{{\sl J. Physics Cond. Mat.} {\bf #1} (#2) #3}
\def\JPSJ#1#2#3{{\sl J. Phys. Soc. Japan} {\bf #1} (#2) #3}
\def\Phy#1#2#3{{\sl Physica} {\bf #1} (#2) #3}
\def\JSM#1#2#3{{\sl J. Soviet Math.} {\bf #1} (#2) #3}
\def\MPLA#1#2#3{{\sl Mod. Phys. Lett.} {\bf A#1} (#2) #3}
\def\MPLB#1#2#3{{\sl Mod. Phys. Lett.} {\bf B#1} (#2) #3}
\def\JETP#1#2#3{{\sl Sov. Phys. JETP} {\bf #1} (#2) #3}
\def\JETPL#1#2#3{{\sl Sov. Phys. JETP Lett.} {\bf #1} (#2) #3}
\def\SPD#1#2#3{{\sl Sov. Phys. Dokl.} {\bf #1} (#2) #3}
\def\CMH#1#2#3{{\sl Comment. Math. Helv.} {\bf #1} (#2) #3}
\def\ZP#1#2#3{{\sl Z.Phys.} {\bf #1} (#2) #3}
\def\ZPB#1#2#3{{\sl Z.Phys.} {\bf B#1} (#2) #3}
\def\sci#1#2#3{{\sl Science} {\bf #1} (#2) #3}
\def\LNC#1#2#3{{\sl Lett. Nuovo Cimento} {\bf #1} (#2) #3}
\def\HPA#1#2#3{{\sl Helv. Phys. Acta} {\bf #1} (#2) #3}
\def\SSC#1#2#3{{\sl Solid State Comm.} {\bf #1} (#2) #3}

\def\sgn{{\rm sgn}}
\def\p{p^{^{\!\!\!\circ}}}
\def\q{q^{^{\!\!\!\circ}}}
\def\dv{|0)}
\def\ddv{(0|}
\def\nddv{{(\tilde 0|}}
\def\qed{{\smallfonts\bf q.e.d.}}
\def\vac{|0\rangle}
\def\dvac{\langle 0|}
\def\dbvac{\bar{\langle 0|}}
\def\bvac{\bar{\vac}}
\def\up{\uparrow}
\def\half{{1\over 2}}
\def\l{\lambda}
\def\m{\mu}
\def\eps{\epsilon}
\def\half{{1\over 2}}
\def\frac#1#2{{#1\over #2}}
\def\B#1{{\cal B}(#1)}
\def\D#1{{\cal D}(#1)}
\def\C#1{{\cal C}(#1)}
\def\A#1{{\cal A}(#1)}
\def\Idoubled#1{{\rm I\kern-.22em #1}}
\def\Odoubled#1{{\setbox0=\hbox{\rm#1}%
     \dimen@=\ht0 \dimen@ii=.04em \advance\dimen@ by-\dimen@ii
     \rlap{\kern.26em \vrule height\dimen@ depth-\dimen@ii width.075em}\box0}}
\def\Real{\Idoubled R}
\def\Complex{\Odoubled C}
\def\xib{\bar{\xi}}

\centerline{\tenbf DUAL FIELD APPROACH TO CORRELATION FUNCTIONS}
\baselineskip=16pt
\centerline{\tenbf IN THE HEISENBERG XXZ SPIN CHAIN}
\vspace{0.8cm}
\centerline{\tenrm FABIAN H.L. ESSLER}
\baselineskip=13pt
\centerline{\tenit Physikalisches Institut der Universit\"at Bonn,
Nussallee 12}
\baselineskip=12pt
\centerline{\tenit D-53115 Bonn, GERMANY}
\vspace{0.3cm}
\centerline{\tenrm and}
\vspace{0.3cm}
\centerline{\tenrm VLADIMIR E. KOREPIN}
\baselineskip=13pt
\centerline{\tenit Institute for Theoretical Physics}
\baselineskip=12pt
\centerline{\tenit State University of New York at Stony Brook}
\centerline{\tenit Stony Brook, NY 11794-3840, USA}
\vspace{0.9cm}
\centerline{\nineit to appear in the proceedings of {\tenit SMQFT}, Los
Angeles 1994}
\vspace{0.9cm}
\abstracts{We study zero temperature correlation functions
of the spin-$1\over 2$ Heisenberg XXZ model
in the critical regime $-1< \Delta\leq 1$ in a magnetic field by means
of the {\tenit Dual Field Approach}. We show for one particular
example how to derive determinant representations for correlation
functions and how to use these to embed the correlation functions in
integrable systems of integro-difference equations (IDE).
These IDE are associated with a Riemann-Hilbert problem.}

\vfil
\rm\baselineskip=14pt
\section{Introduction}
The evaluation of correlation functions in integrable one-dimensional
quantum systems is one of the main outstanding problems in
Mathematical Physics. Quite recently there has been exciting progress
in this direction: the Kyoto group succeeded in deriving
integral representations for some correlation functions of the Heisenberg
XXZ model in the massive regime $\Delta >1$ by taking advantage of the
infinite quantum affine symmetry of the model on the infinite
chain\citeup{rims1,rims2}. These integral representations are most powerful
for studying the {\sl short distance} behaviour of correlators,
whereas it is not obvious how to extract the large distance asymptotics.
Also it is not straightforward to extend this approach to the critical
regime $-1< \Delta < 1$ or to include an external magnetic field.\\
These issues can be very naturally addressed in the
framework of the {\sl Dual Field Approach} (DFA) to correlation functions,
which was pioneered in \hskip-1pt\citelow{iik1,iiks,iikv,k1,k2,ks1,ks2} for
the example of the $\delta$-function Bose gas\citeup{ll,lieb}.
A detailed and complete exhibition of this work can be found in the
book \citelow{vladb}.
By means of the DFA it is possible to derive determinant representations
for correlation functions in practically any integrable model, for which
the Algebraic Bethe Ansatz can be formulated. Using the determinant
representation one can then obtain explicit expressions for the
{\sl large distance} asymptotics of correlation functions (even at finite
temperature), and the inclusion of an external magnetic field poses no
problem. The DFA thus nicely complements the approach of the Kyoto group.
Here we will apply the DFA to the Heisenberg XXZ chain at zero temperature
in a magnetic field $h$, {\sl i.e.} the hamiltonian (with periodic boundary
conditions)
\begin{equation}
{H=\sum_{j=1}^L \sigma^x_j\sigma^x_{j+1}+\sigma^y_j\sigma^y_{j+1}
+\Delta \left(\sigma^z_j\sigma^z_{j+1}-1\right) - h\sum_{j=1}^L
\sigma_j^z\ ,\ -1<\Delta\leq 1, }
\label{H}
\end{equation}
where $\sigma^z=\left(\matrix{1&0\cr 0&-1\cr}\right)$,
$\sigma^x=\left(\matrix{0&1\cr 1&0\cr}\right)$,
$\sigma^y=\left(\matrix{0&-i\cr i&0\cr}\right)$, and
$\Delta=\cos(2\eta),\ {\pi\over 2}<\eta\leq \pi$.
\vskip .5cm\noindent

Our discussion below follows the logic of the DFA:
in section 2 we recall the formulation of the Algebraic
Bethe Ansatz (ABA) for the XXZ chain,  and then use the
ABA to express correlation functions in terms of determinants of
Fredholm integral operators. In section 3 we discuss how to
embed these determinants in systems of integrable integro-difference
equations (IDE) (this is analogous to describing quantum correlation
functions by means of differential equations\citeup{barouch,tracy}).

\section{Determinant Representations}

Let us first review the main features of the Algebraic Bethe Ansatz (ABA)
for the XXZ Heisenberg magnet\citeup{ft2,krs,stf,ft1,takh}.
Starting point and central object of the Quantum Inverse Scattering Method
(see {\sl e.g.} \citelow{ks,takh}) is the R-matrix, which is a solution of
the Yang-Baxter equation\citeup{ybe,baxter2}. For the case of the XXZ model
it is of the form
\begin{equation}
{R(\lambda ,\mu) =\left(\matrix{f(\mu ,\lambda) &0&0&0\cr
0&g(\mu ,\lambda)&1&0\cr 0&1&g(\mu ,\lambda)&0\cr 0&0&0&f(\mu ,\lambda)\cr
}\right) ,}
\label{R}
\end{equation}
where
\begin{equation}
{f(\lambda ,\mu) = {\sinh(\lambda -\mu +2i\eta)\over
\sinh(\lambda -\mu)} \ ,\qquad g(\lambda ,\mu) = {i\sin(2\eta)\over
\sinh(\lambda -\mu)} \ .}
\label{fgxxz}
\end{equation}
The R-matrix is a linear operator on the tensor product of two
two-dimensional linear spaces: $R(\mu)\in
End(\Complex^2\otimes\Complex^2)$.  {}From the R-matrix (\ref{R}) one
can construct an $L$-operator of a "fundamental spin model" (see {\sl
e.g.} \citelow{vladb} p.126) by considering $R(\mu)\Pi$, (where
$\Pi$ is the permutation matrix on $\Complex^2\otimes\Complex^2$)
as an operator-valued matrix by identifying one of the linear spaces with
the two-dimensional Hilbert space ${\cal H}_n$ of $SU(2)$-spins over
the n'th site of a lattice of length $L$
\begin{equation}
L_n(\mu) =\left(\matrix{\sinh(\mu -i\eta\sigma_n^z)
&-i\sin(2\eta) \sigma_n^-\cr -i\sin(2\eta) \sigma_n^+
&\sinh(\mu +i\eta\sigma_n^z)\cr}\right)\ .
\label{lop}
\end{equation}
The Yang-Baxter equation for $R$ implies the following relations for
the $L$-operator
\begin{equation}
{R(\lambda -\mu) \left(L_n(\lambda)\otimes L_n(\mu)\right)
=  \left(L_n(\mu)\otimes L_n(\lambda)\right) R(\lambda -\mu) \ .}
\label{intL}
\end{equation}
{}From the ultralocal $L$-operator the monodromy matrix is constructed as
\begin{equation}
{T(\mu) = L_L(\mu)L_{L-1}(\mu)\ldots L_1(\mu) =
\left(\matrix{A(\mu)&B(\mu)\cr C(\mu)&D(\mu)\cr}\right) \ .}
\label{monodr}
\end{equation}
Eq. (\ref{intL}) can be lifted to the level of the monodromy
matrix
\begin{equation}
{R(\lambda -\mu) \left(T(\lambda)\otimes T(\mu)\right)
=  \left(T(\mu)\otimes T(\lambda)\right) R(\lambda -\mu) \ .}
\label{intT}
\end{equation}
Below we will repeatedly use especially the following matrix elements
of (\ref{intT})
\begin{eqnarray}
[B(\l),B(\m)]&=&0=[C(\l),C(\m)]\nonumber\cr
[B(\l),C(\m)]&=& g(\l ,\m) \left(D(\l) A(\m) - D(\m) A(\l)
\right)\nonumber \cr
D(\m)B(\l)&=& f(\l ,\m) B(\l)D(\m) + g(\m ,\l)B(\m)D(\l) \nonumber \cr
A(\m)B(\l)&=& f(\m ,\l) B(\l)A(\m) + g(\l ,\m)B(\m)A(\l) .
\label{intTme}
\end{eqnarray}
By tracing (\ref{intT}) over the matrix space one obtains a
one-parameter family of commuting {\sl transfer matrices}
$\tau(\mu)=tr(T(\mu))=A(\mu)+D(\mu)$: $[\tau(\mu),\tau(\nu)]=0$.
The transfer matrix is the generating functional of an infinite number
of mutually commuting conserved quantum operators (via expansion in
powers of spectral parameter). One of these operators is the hamiltonian
\begin{equation}
{H = -2i\sin(2\eta) {\partial\over \partial\mu}
\ln(\tau(\mu))\bigg|_{\mu=-i\eta} - 2L\cos(2\eta) -2hS^z\ .}
\label{Hxxz}
\end{equation}
Below we also make use of some properties of {\sl inhomogenous} XXZ
models, which are constructed in the following way: we first note that the
intertwining relation for the $L$-operator (\ref{intL}) still
holds, if we shift both spectral parameters $\l$ and $\mu$ by an
arbitrary amount $\nu_n$, {\sl i.e.}
\begin{equation}
{R(\lambda -\mu) \left(L_n(\lambda -\nu_n)\otimes
L_n(\mu -\nu_n)\right) =  \left(L_n(\mu -\nu_n)\otimes L_n(\lambda
-\nu_n)\right) R(\lambda -\mu) \ .}
\label{intL2}
\end{equation}
The reason for this is of course that the $R$-matrix only depends on
the difference of spectral parameters. We now can construct a monodromy
matrix as
\begin{equation}
{T_{inh}(\l) = L_L(\l -\nu_L)L_{L-1}(\l
-\nu_{L-1})\ldots L_1(\l -\nu_1) = \left(\matrix{\A{\mu}&\B{\mu}\cr
\C{\mu}&\D{\mu}\cr}\right) \ .}
   \label{Tinhom}
\end{equation}
The inhomogeneous monodromy matrix (\ref{Tinhom}) obeys the same
intertwining relation (\ref{intT}) as (\ref{monodr}).

The ABA deals with the construction of simultaneous eigenstates of the
transfer matrix and the hamiltonian. Starting point is the choice of a
{\sl reference state}, which is a trivial eigenstate of $\tau(\mu)$. In
our case we make the choice $\vac = \otimes_{n=1}^L |\up\rangle_n$,
{\sl i.e.} we choose the completely ferromagnetic state. The action of
the $L$-operator (\ref{lop}) on $|\up\rangle_n$ can be easily computed
and implies the following actions of the matrix elements of the
monodromy matrix
\begin{eqnarray}
A(\mu)\vac &=& a(\mu)\vac\ ,\quad a(\mu) =
\left(\sinh(\mu -i\eta)\right)^L\ ,\nonumber\cr
D(\mu)\vac &=& d(\mu)\vac\ ,\quad d(\mu) =
\left(\sinh(\mu +i\eta)\right)^L,\ \nonumber\cr
C(\mu)\vac &=&0\ ,\nonumber\cr
B(\mu)\vac &\neq0\ ,
   \label{ADxxz}
\end{eqnarray}
{}From (\ref{ADxxz}) it follows that $B(\lambda)$ plays the role of a
creation operator, {\sl i.e.} one can construct a set of states of the
form
\begin{equation}
{\Psi_N(\lambda_1 ,\ldots ,\lambda_N) = \prod_{j=1}^N
B(\lambda_j)\vac\ .}
   \label{states}
\end{equation}
The requirement that the states (\ref{states}) ought to be eigenstates
of the transfer matrix $\tau(\mu)$ puts constraints on the allowed values
of the parameters $\lambda_n$: the set $\{\lambda_j\}$ must be a solution
of the following system of coupled algebraic equations, called {\sl Bethe
equations}\citeup{bethe,orbach}
\begin{equation}
{{a(\lambda_j)\over d(\lambda_j)} =
\prod_{\scriptstyle k=1\atop \scriptstyle k\neq j}^N
{f(\lambda_k,\lambda_j)\over f(\lambda_j,\lambda_k)}\ , j=1,\ldots ,N\ .}
   \label{bae}
\end{equation}
These equations are the basis for studying ground state, excitation
spectrum and thermodynamics of Bethe Ansatz solvable models. For the case
of the XXZ model with $\Delta > -1$ (the case we are interested in
here) it was proved by C.N. Yang and C.P. Yang\citeup{yaya1,yaya2}
that the ground state is characterized by a set of {\sl real}
$\lambda_j$ subject to the Bethe equations (\ref{bae}). Without an
external magnetic field ($h=0$) their number is $N=L/2$. In the
thermodynamic limit the ground state is described by means of an
integral equation for the density of spectral parameters
$\rho(\lambda)$\citeup{hulthen,yaya2}
\begin{equation}
{2\pi \rho(\lambda) - \int_{-\Lambda}^{\Lambda} d\mu\
K(\lambda ,\mu)\ \rho(\mu)\ = D(\l)\ ,}
   \label{gsie}
\end{equation}
where the integral kernel $K$ and the driving term $D$ are given by
\begin{equation}
K (\m, \l ) = \frac{\sin(4\eta)}{\sinh(\mu
- \lambda +2i\eta)\sinh(\mu - \lambda -2i\eta)}\ ,\ D(\l) =
{-\sin(2\eta)\over \sinh(\lambda -i\eta) \sinh(\lambda +i\eta) }\ .
   \label{kernel}
\end{equation}
Here $\Lambda$ depends on the external magnetic field $h$.  The physical
picture of the ground state is that of a filled Fermi sea with boundaries
$\pm \Lambda$. The dressed energy of a particle in the sea is given by the
solution of the integral equation\citeup{taka1,taka2}
\begin{equation}
{\eps(\lambda) - {1\over
2\pi}\int_{-\Lambda}^{\Lambda} d\mu\ K(\lambda ,\mu)\ \eps(\mu)\ =
2h -{2 (\sin(2\eta))^2\over \sinh(\lambda -i\eta) \sinh(\lambda
+i\eta) }\ .}
   \label{gsie2}
\end{equation}
The requirement of the vanishing of the dressed energy at the Fermi
boundary $\eps(\pm\Lambda)=0$ determines the dependence of $\Lambda$ on
$h$. For small $h$ this relation can be found explicitly by means of a
Wiener-Hopf analysis\citeup{yaya2}.
For $h\geq h_c=(2\cos\eta)^2$ the system is in the saturated
ferromagnetic state, which corresponds to $\Lambda=0$.
\newpage
{\subsection{Two-Site Generalized Model}}
For the evaluation of correlation functions the so-called ``two-site
generalized model'' has proven a useful tool. {}From the
mathematical point of view this is simply the application of the
co-product associated with the algebra defined by (\ref{intT}). The
main idea is to divide the chain of length $L$ into two parts and
associate a monodromy matrix with both sub-chains, {\sl i.e.}
\begin{equation}
{T(\mu) = T(2,\mu)T(1,\mu)\ ,\
T(i,\mu) =\left(\matrix{A_i(\mu) &B_i(\mu)\cr C_i(\mu)
&D_i(\mu)\cr}\right)\ (i=1,2)\ .}
\label{t12}
\end{equation}
In terms of $L$-operators the monodromy matrices are given by
\begin{eqnarray}
T(2,\mu) &=& L_L(\mu)L_{L-1}(\mu)\ldots
L_n(\mu)\ ,\cr
T(1,\mu) &=& L_{n-1}(\mu)L_{n-2}(\mu)\ldots L_1(\mu)\ .
\label{t12l}
\end{eqnarray}
By construction it is clear that both monodromy matrices $T(i,\mu)$
fulfill the same intertwining relation (\ref{intT}) as the complete
monodromy matrix $T(\mu)$.
Similarly the reference state for the complete chain is decomposed
into a direct product of reference states $\vac_i$ for the two
sub-chains $\vac = \vac_2\otimes\vac_1$. The resulting structure can
be summarized as
\begin{eqnarray}
A_i(\mu)\vac_i &=& a_i(\mu)\vac_i\ ,\quad
D_i(\mu)\vac_i = d_i(\mu)\vac_i\ ,\cr
C(\mu)\vac_i &=&0\ ,\quad B_i(\mu)\vac_i\neq 0\ ,
\label{abcdi}
\end{eqnarray}
where the eigenvalues $a$ and $d$ in (\ref{ADxxz}) are given by
$a(\mu) =a_2(\mu)a_1(\mu)$ and $d(\mu) =d_2(\mu)d_1(\mu)$.
The creation operators $B(\mu)$ for the complete chain are decomposed
as $B(\mu) = A_2(\mu)\otimes B_1(\mu) +B_2(\mu) \otimes D_1(\mu)$,
which implies that eigenstates of the transfer matrix can be
represented as
\begin{eqnarray}
\prod_{j=1}^N B(\lambda_j)\vac &=&
\sum_{I,II}\prod_{j\in I}^{n_1}\prod_{k\in II}^{n_2} a_2(\lambda_j^I)
d_1(\lambda_k^{II})\cr
&&\quad \times f(\lambda_j^I,\lambda_k^{II})
\left(B_2(\lambda_k^{II})\vac_2\right)\otimes
\left(B_1(\lambda_j^{I})\vac_1\right)\ ,
\label{bbb}
\end{eqnarray}
where the sum is over all partitions
$\{\lambda_j^I\}\cup\{\lambda_k^{II}\}$ of the set $\{\lambda_j\}$ with
${\rm card}\{\lambda^I\}=n_1$, ${\rm card}\{\lambda^{II}\}=n_2=N-n_1$.  A
similar equation holds for dual states
\begin{eqnarray}
\langle 0|\prod_{j=1}^N C(\lambda_j) &=&
\sum_{I,II}\prod_{j\in I}^{n_1}\prod_{k\in II}^{n_2} d_2(\lambda_j^I)
a_1(\lambda_k^{II})\cr
&&\quad \times f(\lambda_k^{II},\lambda_j^{I})
\left({_1}\dvac C_1(\lambda_j^{I})\right)\otimes
\left({_2}\dvac C_2(\lambda_k^{II})\right)\ .
\label{ccc}
\end{eqnarray}

\newpage
{\subsection{Reduction of Correlators to Scalar Products}}
We are now in the position express correlation functions in the
framework of the ABA. We will concentrate on the Ferromegnetic String
Formation Probability (FSFP) correlation function, which is defined as
follows
\begin{equation}
P(m) = \langle GS |\ \prod_{j=1}^m P_j \ | GS \rangle ,
\label{px}
\end{equation}
where $|GS\rangle$ is the antiferromagnetic ground state and
$P_j = {1\over2} (\sigma_j^z + 1)$ is the projection operator onto the
state with spin up at site $j$. The physical meaning of $P(m)$
is the probability of finding a ferromagnetic string ($m$ adjacent
parallel spins up) in the ground state $|GS\rangle$ of the model
(\ref{H}) for a given value of the magnetic field $h$. From a
technical point of view this correlator turns out the simplest one
in the DFA\citeup{vladb}. Let us now rewrite (\ref{px}) in
terms of the ABA. Noting that ${1\over2} (\sigma_j^z +1)=\exp\left(
{\alpha\over 2}(1-\sigma_j^z)\right)\bigg|_{\alpha=-\infty}$ we find
\begin{equation}
P(m)=\frac{\langle 0|\prod_{j=1}^{{N}}C(\lambda_j)
\exp(\alpha\sum_{l=1}^m\sigma_l^-\sigma_l^+)\prod_{k=1}^{{N}}
B(\lambda_k)|0\rangle}{\langle
0|\prod_{j=1}^{{N}}C(\lambda_j)\prod_{k=1}^{{N}}B(\lambda_k)|0\rangle}
\bigg|_{\alpha =-\infty}\ ,
\label{genfu}
\end{equation}
where the spectral parameters $\lambda_j$ are subject the Bethe
equations (\ref{bae}) for the ground state.
This expression can be rewritten in terms of scalar products by using
the two-site generalized model introduced above: we take the first
sub-chain to contain sites $1$ to $m$ and the second one sites $m+1$
to $L$. We note that due to the projection ($\alpha=-\infty$) in
(\ref{genfu}) the partitions in (\ref{bbb}) and (\ref{ccc}) must be
such that there are no down-spins on the first sub-chain (only the
partition where $n_1=0$ survives), and thus
\begin{equation}
P(m) = {1\over \sigma_{N}}\ {_2}\dvac\prod_{j=1}^N C_2(\l_{j})
\prod_{k=1}^N B_2(\l_{k})\vac_2\ \ \prod_{l=1}^N
a_1(\l_{l})d_1(\l_{l}) ,
\label{genfu2}
\end{equation}
where
\begin{equation}
\sigma_{N} = {\langle 0| \prod_{j=1}^{{N}} C(\lambda_j)
\prod_{k=1}^{{N}}B(\lambda_k)|0\rangle}\ .
\label{Sn}
\end{equation}

\subsection{Determinant Representations for Scalar Products}

In (\ref{genfu2}) we have reduced the computation of $P(m)$ to the
computation of a scalar product of the form
\begin{equation}
{S}_N = {\langle 0| \prod_{j=1}^{N} C(\lambda_j^C)
\prod_{k=1}^{N}B(\lambda_k^B)|0\rangle}\ .
\label{scpr}
\end{equation}
Following \citelow{k2,k3} we will show how to represent (\ref{scpr})
as a determinant.
Here we {\sl do not} assume that the sets of spectral parameters
$\{\l^B\}$ and $\{\l^C\}$ are the same, and we also {\sl do not}
impose the Bethe equations (\ref{bae}).
{}From (\ref{intTme}) and (\ref{ADxxz}) it follows
that scalar products can be represented as\citeup{k1}
\begin{equation}
{{S}_N = \sum_{A,D} \prod_{j=1}^N a(\l_j^A)\prod_{k=1}^N
d(\l_k^D) K_N\left(\matrix{
\{\l^C\}&\{\l^B\}\cr\{\l^A\}&\{\l^D\}\cr}\right)\ ,}
\label{Ks}
\end{equation}
where the sum is over all partitions of $\{\l^C\}\cup \{\l^B\}$ into
two sets $\{\l^A\}$ and $\{\l^D\}$. The coefficients $K_N$ are
functions of the $\l_j$ and are completely determined by the
intertwining relation (\ref{intT}). In particular the $K_N$'s are
identical for the homogeneous model (\ref{monodr}) and the {\sl
inhomogeneous} model (\ref{Tinhom}), {\sl i.e.} the $K_N$'s are
independent of the inhomogeneities $\{\nu_n\}$ and also do not depend
on the lattice length $L$ as long as $N<L$. The reason for this is
that the intertwining relations for the matrix elements ${\cal
A}(\mu)$, ${\cal B}(\mu)$ , ${\cal C}(\mu)$ and ${\cal D}(\mu)$ of
(\ref{Tinhom}) are the same as the ones for the matrix elements
$A(\mu)$, $B(\mu)$, $C(\mu)$, $D(\mu)$ of(\ref{monodr}) (see
above). We will exploit this fact
by considering special inhomogeneous models for which all terms but one
in the sum in (\ref{Ks}) vanish, and then represent this term as a
determinant. The basic tool for representing scalar products as
determinants is a Theorem due to Izergin, Coker and
Korepin\citeup{i1,ick}, which deals with determinant representations for
the {\sl partition functions} of inhomogeneous XXZ models
constructed according to (\ref{Tinhom}):

\vskip .3cm
\noindent{\bf Theorem:} \sl Consider an inhomogeneous XXZ chain of
even length $N$ with inhomogeneities $\nu_j,\ j=1\ldots N$. Let $\vac$
and ${\bar{\vac}}$ be the ferromagnetic reference states with all spins
up and down respectively. Let $\B{\mu}$ and $\C{\mu}$ be the
  creation/annihilation operators over the reference state $\vac$.
  Then the following determinant representation holds
\begin{eqnarray}
{\bar{\dvac}} \prod_{j=1}^N \B{\lambda_j}{\vac} &=&
{\dvac} \prod_{j=1}^N \C{\lambda_j}{\bar{\vac}} \cr
&=& (-1)^N\prod_{\alpha =1}^N \prod_{k =1}^N \sinh(\lambda_\alpha
-\nu_k-i\eta)\ \sinh(\lambda_\alpha -\nu_k+i\eta) \cr
\times\hskip-6pt&&\left(\prod_{1\leq\alpha <\beta\leq N}\!\!
\sinh(\l_\alpha -\l_\beta)\prod_{1\leq k <l\leq N}\!\! \sinh(\nu_l
-\nu_k) \right)^{-1}{\rm det}({\cal M}) \ ,
\label{detZ1}
\end{eqnarray}
where
\begin{equation}
{\cal M}_{\alpha k} = {i\sin(2\eta)\over
\sinh(\l_\alpha -\nu_k -i\eta)\sinh(\l_\alpha -\nu_k +i\eta)}\ .
\label{detZ2}
\end{equation}
\vskip .5cm\rm
Let us now derive explicit expressions for the coefficients $K_N$. It
will be convenient to work with the following sets of spectral
parameters
\begin{eqnarray}
\{\l^{AC}\} &=& \{\l^A\}\cap \{\l^{C}\},\
\{\l^{DC}\} = \{\l^D\}\cap \{\l^{C}\},\cr
\{\l^{AB}\} &=& \{\l^A\}\cap \{\l^{B}\},\
\{\l^{DB}\} = \{\l^D\}\cap \{\l^{B}\},
\end{eqnarray}
with cardinalities
\begin{eqnarray}
n&=&{\rm card}\{\l^{DC}\}={\rm card}\{\l^{AB}\}\ ,\cr
N-n&=&{\rm card}\{\l^{AC}\}={\rm card}\{\l^{DB}\}\ .
\end{eqnarray}
The partition with $n=0$ is characterized by $\{\l^{AC}\}=\{\l^{C}\}$,
$\{\l^{DB}\}=\{\l^{B}\}$, $\{\l^{AB}\}=\emptyset = \{\l^{DC}\}$. The
corresponding coefficient $K_N\left(\matrix{
  \{\l^C\}&\{\l^B\}\cr\{\l^C\}&\{\l^B\}\cr}\right)$ is called {\sl
  highest coefficient}. For highest coefficients the following
determinant representation holds
\begin{eqnarray}
K_N\left(\matrix{\{\l^C\}&\{\l^B\}\cr\{\l^C\}&\{\l^B\}\cr}\right) &=&
\left(\prod_{j>k}g(\l_j^B, \l_k^B)g(\l_k^C, \l_j^C)\right) \prod_{j,k}
h(\l_j^C,\l_k^B) {\rm det}(M^{B}_{C})\, \cr
h(\l,\m) &=& {f(\l,\m)\over g(\l,\m)} = {\sinh(\l-\m+2i\eta)\over
i\sin(2\eta)} ,\cr
\left(M^B_C\right)_{jk} =t(\l_j^C,\l^B_k) &=& {g(\l_j^C,\l^B_k)\over
h(\l_j^C,\l^B_k)} = {(i\sin(2\eta))^2\over
\sinh(\l_j^C-\l_k^B+2i\eta)\sinh(\l_j^C-\l_k^B)}\ .
\label{Kcbcb}
\end{eqnarray}
The proof of (\ref{Kcbcb}) is as follows: Consider an inhomogeneous
XXZ model on a lattice of length $N$ with inhomogeneities $\nu_j =
\l_j^C+i\eta$. We have $a(\l) = \prod_{j=1}^N\sinh(\l-\l_j^C-2i\eta)$
and $d(\l) = \prod_{j=1}^N\sinh(\l-\l_j^C)$. Inspection of (\ref{Ks})
yields that in this situation only one term in the sum of the r.h.s of
(\ref{Ks}) survives, namely the one with $\{\l^D\} = \{\l^B\}$.  Thus
for this special scalar product we obtain
\begin{equation}
S_N\bigg|_{\nu_j = \l^C_j+i\eta} = K_N\left(\matrix{
\{\l^C\}&\{\l^B\}\cr\{\l^C\}&\{\l^B\}\cr}\right) \prod_{j,k}
\sinh(\l_j^C-\l_k^C-2i\eta)\prod_{m,l}\sinh(\l_m^B-\l_l^C)\ .
\label{hcoeff}
\end{equation}
On the other hand $B(\l)$ flips one spin, and as we have chosen $N$ to
be the length of the lattice we find that $\prod_{j=1}^NB(\l_j)\vac$
is proportional to the ferromagnetic state with all spins flipped, and
thus orthogonal to all states in a basis other than ${\bar{\vac}}$.
Thus
\begin{equation}
S_N\bigg|_{\nu_j = \l^C_j+i\eta} = {\langle 0| \prod_{j=1}^{N}
C(\lambda_j^C)\bvac \dbvac \prod_{k=1}^{N}B(\lambda_k^B)|0\rangle}.
\end{equation}
By the Theorem both factors can be represented as determinants. By direct
computation we find for one of the factors
\begin{equation}
\langle 0| \prod_{j=1}^{N}   C(\lambda_j^C)\bvac  =
\prod_{j,k}\sinh(\l_k^C -\l_j^C-2i\eta) \ .
\end{equation}
Using the determinant representation given by the Theorem on the other
factor we arrive at (\ref{Kcbcb}).\\
Arbitrary coefficients $K_N$ are expressed in terms of highest
coefficients as follows
\begin{eqnarray}
K_N\left(\matrix{
\{\l^C\}&\{\l^B\}\cr\{\l^A\}&\{\l^D\}\cr}\right) &=& \left(\prod_{j\in
AC}\prod_{k\in DC} f(\l_j^{AC}, \l_k^{DC})\right)\left(\prod_{l\in
AB}\prod_{m\in DB} f(\l_l^{AB}, \l_m^{DB})\right)\cr
\times&& K_n\left(\matrix{
\{\l^{AB}\}&\{\l^{DC}\}\cr\{\l^{AB}\}&\{\l^{DC}\}\cr}\right)
K_{N-n}\left(\matrix{\{\l^{AC}\}&\{\l^{DB}\}\cr\{\l^{AC}\}&
\{\l^{DB}\}\cr}\right) \ .
\label{Kcbad}
\end{eqnarray}
To prove (\ref{Kcbad}) we consider an inhomogeneous XXZ model with
inhomogeneities $\{\nu_j\} = \{\l_j^{AB}+i\eta\}\cup
\{\l_j^{AC}+i\eta\}$. Now only the term proportional to
$K_N\left(\matrix{ \{\l^C\}&\{\l^B\}\cr\{\l^A\}&\{\l^D\}\cr}\right)$
in the sum on the r.h.s of. (\ref{Ks}) survives. Proceeding as above
we arrive at (\ref{Kcbad}).
\vskip .5cm
Combining (\ref{Kcbcb}) and (\ref{Kcbad}) with (\ref{Ks}) we obtain
the following expression for general scalar products of XXZ magnets
\begin{eqnarray}
S_N &=& \prod_{j>k}
g(\l_j^C,\l_k^C)g(\l_k^B,\l_j^B) \sum_{}{\rm sgn}(P_C){\rm
sgn}(P_B)\prod_{j,k} h(\l_j^{AB},\l_k^{DC})
\prod_{l,m} h(\l_l^{AC},\l_m^{DB})\cr
&&\times\ \prod_{l,k} h(\l_l^{AC},\l_k^{DC})\prod_{j,m}h(\l_j^{AB},\l_m^{DB})
{\rm det}(M^{AB}_{DC}){\rm det}(M^{AC}_{DB})\ ,
\label{Sn2}
\end{eqnarray}
where $P_C$ is the permutation $\{ \l_1^{AC},\ldots ,
\l_n^{AC},\l_1^{DC},\ldots ,\l^{DC}_{N-n}\}$ of $\{\l_1^C,\ldots
,\l_N^C\}$, $P_B$ is the permutation $\{ \l_1^{DB},\ldots ,
\l_n^{DB},\l_1^{AB},\ldots ,\l^{AB}_{N-n}\}$ of $\{\l_1^B,\ldots
,\l_N^B\}$, $sgn(P)$ is the sign of the permutation $P$, and
\begin{equation}
 \left(M^{AB}_{DC}\right)_{jk} = t(\l_j^{AB},\l_k^{DC})
  d(\l_k^{DC})a(\l_j^{AB}) ,\quad t(\l,\m) = {(g(\l,\m))^2\over
    f(\l,\m)}\ .
\end{equation}

The most important step in the DFA follows next: we introduce {\sl
dual quantum fields} in order to simplify (\ref{Sn2}) and obtain
a manageable expression for scalar products. This step was first
carried out for the delta-function Bose gas\citeup{k2}. The
XXZ case can be treated very similarly, so that we will be brief in
our discussion.
The fundamental observation is that the r.h.s. in(\ref{Sn2}) looks
like the determinant of the {\sl sum} of two matrices, {\sl i.e.}
let $A$ and $B$ be two $N\times N$ matrices over $\Complex$. Then the
determinant of their sum can be decomposed as follows
\begin{equation}
{\rm det}(A+B) = \sum_{P_r,P_c}{\rm sgn}(P_r){\rm
sgn}(P_c) {\rm det}(A_{P_rP_c}){\rm det}(B_{P_rP_c})\ .
\label{detAB}
\end{equation}
Here $P_r$ and $P_c$ are partitions of the $N$ rows and columns into
two subsets ${\cal R}$, ${\bar{\cal R}}$ and ${\cal C}$ $\bar{\cal C}$
of cardinalities $n$ (for ${\cal R}$, ${\cal C}$) and $N-n$ (for
$\bar{\cal R}$, $\bar{\cal C}$) respectively, $A_{P_rP_c}$ is the
$n\times n$ matrix obtained from $A$ by removing all $\bar{\cal
R}$-rows and $\bar{\cal C}$-columns,  and $B_{P_rP_c}$ is the
$N-n\times N-n$ matrix obtained from $B$ by removing all ${\cal
R}$-rows and ${\cal C}$-columns. Finally ${\rm sgn}(P_r)$ is the
parity of the permutation obtained from $(1,\ldots ,N)$ by moving all
${\cal R}$-rows to the front.
Comparison of (\ref{Sn2}) with (\ref{detAB}) shows that one does not get the
$h(\l,\m)$-factors by simply taking the determinant of the sum of the
matrices $M_{jk}$. This leads to the introduction of the auxiliary
variables $\Phi_A(\l)$ and $\Phi_D(\l)$, which are called {\sl dual quantum
fields} and which are represented as sums of
``momenta'' $P_A$ and ``coordinates'' $Q_A$ as follows
\begin{eqnarray}
\Phi_A(\l) &=& Q_A(\l) + P_D(\l),\quad
\Phi_D(\l) = Q_D(\l) + P_A(\l),\cr
[P_D(\l), Q_D(\m)]&=& \ln(h(\l,\m)),\quad [P_A(\l), Q_A(\m)]=
\ln(h(\m,\l))\ .
\label{DQF}
\end{eqnarray}
All other commutators of $P$'s and $Q$'s vanish. A very important
property of the fields $\Phi$ is that they commute for different
values of spectral parameters
\begin{equation}
[\Phi_A(\l), \Phi_D(\m)] = 0 =[\Phi_A(\l), \Phi_A(\m)]
=[\Phi_D(\l), \Phi_D(\m)]\ .
\end{equation}
The dual quantum fields act on a bosonic Fock space with reference
states $\dv$ and $\ddv$ defined by
\begin{equation}
\label{dfs}
P_a(\l)\dv =0\ ,\quad \ddv Q_a(\l) =0\ ,\ a=A,D\ , \ddv
0)=1\ .
\end{equation}
Using the dual fields it is now possible to recast (\ref{Sn2}) as a
determinant of the sum of two matrices in the following way
\begin{eqnarray}
S_N &=& \prod_{j>k} g(\l_j^C,\l_k^C)
g(\l_k^B,\l_j^B) \ddv \det S\dv\ ,\cr
S_{jk} &=& t(\l_j^C,\l_k^B) a(\l_j^C)d(\l_k^B)
\exp\left(\Phi_A(\l_j^C)+\Phi_D(\l_k^B) \right)\cr
&&\qquad + t(\l_k^B,\l_j^C) d(\l_j^C)a(\l_k^B)
\exp\left(\Phi_D(\l_j^C)+\Phi_A(\l_k^B) \right)\ .
\label{Sn3}
\end{eqnarray}
It is possible to further simplify (\ref{Sn3}) by eliminating one
dual field: we define a new dual vacuum $\nddv$ according to
\begin{equation}
\nddv = \ddv \exp\left(\sum_{j=1}^N
P_D(\l_j^C)+P_A(\l_j^B)\right) \ ,\ \nddv 0)=1\ ,
\label{ndv}
\end{equation}
and a new dual field
\begin{eqnarray}
\varphi(\l) &=& p(\l) + q(\l),\cr
q(\l) &=& Q_A(\l)-Q_D(\l) - \nddv Q_A(\l)-Q_D(\l)\dv,\cr
p(\l) &=& P_D(\l)-P_A(\l) \ ,\ \nddv q(\l) = 0 = p(\l) \dv\ ,\cr
[p(\l), q(\m)]&=& -\ln(h(\l,\m)h(\m,\l))\ ,\cr
[p(\l),p(\m)]&=& 0 =[q(\l),q(\m)]=[\varphi(\l),\varphi(\m)]\ .
\label{phi}
\end{eqnarray}
In terms of this field we obtain the following determinant
representation
\begin{eqnarray}
S_N &=& \prod_{j>k} g(\l_j^C,\l_k^C)
g(\l_k^B,\l_j^B) \prod_{j=1}^N
a(\l_j^C)d(\l_j^B)\prod_{j,k}h(\l_j^C,\l_k^B) \nddv \det S\dv\ ,\cr
S_{jk} &=& t(\l_j^C,\l_k^B) + t(\l_k^B,\l_j^C) {r(\l_k^B)\over r(\l_j^C)}
\exp\left(\varphi(\l_k^B)-\varphi(\l_j^C) \right)\cr
&&\qquad\qquad\times \prod_{m=1}^N {h(\l_k^B,\l_m^B)
h(\l_m^C,\l_j^C)\over h(\l_m^C,\l_k^B) h(\l_j^C,\l_m^B)}
\label{Sn3new}
\end{eqnarray}
where $r(\l) = {a(\l)\over d(\l)}$.

Norms of Bethe wave functions are special cases of (\ref{scpr}).
These were first conjectured in \citelow{gaudin2} (see
also \citelow{gaudin3}). This conjecture was generalized and proved
in \citelow{k1}. The result is
\begin{eqnarray}
&&\langle 0| \prod_{j=1}^{N} C(\lambda_j)
\prod_{k=1}^{N}B(\lambda_k)|0\rangle= \prod_{j\neq k} f(\l_j,\l_k)
\prod_{j=1}^N a(\l_j)d(\l_j) {\det{}\cal N}^\prime\ ,\cr
{\cal N^\prime}_{jk}&=& \sin(2\eta)\left(
       -K(\l_j,\l_k) + i\ \delta_{jk}
      {\partial\over\partial\l_j}\left[\ln(r(\l_j)) +
      \sum_{n=1}^N\ln({h(\l_j,\l_n)\over h(\l_n,\l_j)})\right]\right) \cr
  & =& \sin(2\eta) \left(-K(\l_j,\l_k) + \delta_{jk}
       \left[ i\ {\partial\over\partial\l_j} \ln(r(\l_j)) +
        \sum_{n=1}^N K(\l_j,\l_n) \right] \right) ,
\label{norm3}
\end{eqnarray}
where $K(\l,\m)$ and $h(\l,\m)$ are defined in (\ref{kernel}).
We note that (\ref{norm3}) can be obtained directly from
(\ref{Sn3new}) by setting all dual fields to zero and then taking the
sets of spectral parameters $\{\l^C\}$ and $\{\l^B\}$ equal and
imposing the Bethe equations (\ref{bae}).

\subsection{Determinant Representation for FSFP correlation function}

Let us now use the machinery built up above
express $P(m)$ as a determinant. We will proceed in two steps: we
first will analyse (\ref{genfu2}) {\sl without} using that
$\{\l^B\}=\{\l^C\}=\{\l\}$ and {\sl without} imposing the
Bethe-equations (\ref{bae}). In the second step we will then impose
these two constraints.
Using (\ref{Sn3new}) we can represent the scalar product in the
two-site generalized models in (\ref{genfu2}) as a determinant and
obtain
\begin{eqnarray}
P(m) &=& {1\over \sigma_{N}} \prod_{j>k}g(\l_j^C,\l_k^C)g(\l_k^B,\l_j^B)
\ddv\det s(\{\l^C\},\{\l^B\})\dv \ ,\cr
(s(\{\l^C\},\{\l^B\}))_{jk}&=&\!\!  a_1(\l_j^C) d_1(\l_k^B)
\Bigg[t(\l_j^C,\l_k^B) a_2(\l_j^C)d_2(\l_k^B)
\exp\left(\Phi_{A}(\l_j^C)+\Phi_{D}(\l_k^B) \right)\cr
&&\qquad + t(\l_k^B,\l_j^C) d_2(\l_j^C)a_2(\l_k^B)
\exp\left(\Phi_{D}(\l_j^C)+\Phi_{A}(\l_k^B) \right)\Bigg].
\label{genfu3}
\end{eqnarray}
Here the dual fields are defined according to
\begin{eqnarray}
\Phi_{A}(\l) &=& Q_{A}(\l) +
P_{D}(\l),\quad \Phi_{D}(\l) = Q_{D}(\l) +
P_{A}(\l),\cr
[P_{D}(\l), Q_{D}(\m)]&=&\ln(h(\l,\m)) , \quad [P_{A}(\l), Q_{A}(\m)]=
\ln(h(\m,\l))\ .
\label{df}
\end{eqnarray}
All other commutators vanish. The reference state $\dv$ and its
dual $\ddv$ are defined via
\begin{equation}
P_a(\l)\dv =0\ ,\quad \ddv Q_a(\l) =0\ ,\ a=A,D\ , \ddv 0)=1\ .
\label{df2}
\end{equation}

So far we have not used the fact that we are dealing with expectation
values of Bethe states, {\sl i.e.} we have neither used the fact that
$\{\l^C\}=\{\l^B\}=\{\l\}$ nor imposed the Bethe equations
(\ref{bae}). Let us now impose these constraints. Like for the case
of scalar products one of the dual fields can be eliminated

We define a new dual vacuum $\nddv$ and a new dual field
according to
\begin{eqnarray}
\nddv &=& \ddv \exp\left(\sum_{j=1}^{N}
P_{D}(\l_j)+P_{A}(\l_j)\right) \ ,\ \nddv 0)=1\ ,\cr
\phi(\l)&=& \Phi_{A}(\l)-\Phi_{D}(\l)\ ,
\label{ndf}
\end{eqnarray}
The momenta $\p(\l)$ and coordinates $\q(\l)$ of the dual field
$\phi(\l)$ obey the commutation relations
\begin{equation}
 [\q(\mu), \p(\lambda)] = \ln(h(\lambda ,\mu)h(\mu ,\lambda))
\label{ndf3}
\end{equation}
(all other commutators vanish).
By straighforward rewriting of (\ref{genfu3}) in terms
of the new field and the new dual reference state we obtain
\begin{eqnarray}
P(m) &=& {1\over \sigma_{N}}
\prod_{j>k}f(\l_j,\l_k)f(\l_k,\l_j)\prod_{j=1}^{N} a(\l_j)d(\l_j)\nddv
\det{\bar{\cal M}}\dv\ ,\cr
{\bar{\cal M}}_{jk}&=& t(\l_j,\l_k) + t(\l_k,\l_j) {r_2(\l_k)\over
r_2(\l_j)}\exp\left(\phi(\l_k) - \phi(\l_j)\right).
\label{genfu7}
\end{eqnarray}

Here we have used that
\begin{equation}
   \ddv \exp\left(\sum_{j=1}^{N} \Phi_{A}(\l_j)+\Phi_{D}(\l_j)\right) =
   \prod_{j,k}h(\l_j,\l_k)\nddv\ .
\end{equation}
It is found that whereas ${\p}(\l)\dv =0$, the coordinate
${\q}(\l)$ of $\phi(\l)$ do not annihilate the new dual reference
state $\nddv$. Therefore we ``shift'' $\phi(\l)$ by subtracting
its vacuum expectation value in analogy with (\ref{phi})
\begin{equation}
\varphi(\l) = \phi(\l) -\nddv \phi(\l)\dv = p(\l)+q(\l)\ .
\label{ndf2}
\end{equation}
\indent By construction $p$ and $q$ have the same commutation
relations (\ref{ndf2}) as the momenta/coordinates ${\p}(\l)$
and ${\q}(\l)$ of $\phi(\l)$. Furthermore $p(\l)\dv=0$
and $\nddv q(\l)=0$. The shift is equal to
\begin{equation}
\kappa(\l)=\nddv \phi(\l)\dv = \sum_j\ln\left({h(\l,\l_j)\over
h(\l_j,\l)}\right)\ .
\end{equation}
If we replace $\phi$ in (\ref{genfu7}) by $\varphi$ we pick up
additional factors due to the shifts
\begin{eqnarray}
P(m)&=& {1\over \sigma_{N}}
\prod_{j\neq k}f(\l_j,\l_k)\prod_{j=1}^{N} a(\l_j)d(\l_j)\nddv
\det{{\cal G}}\dv\ ,\cr
{{\cal G}}_{jk}&=& t(\l_j,\l_k) + t(\l_k,\l_j) {r_2(\l_k)\over
r_2(\l_j)}e^{\varphi(\l_k) - \varphi(\l_j)}e^{\kappa(\l_k) -
\kappa(\l_j)} .
\label{genfu8}
\end{eqnarray}
The off-diagonal matrix elements of $G$ can be further simplified by simply
imposing the Bethe equations.  Rewriting the Bethe equations (\ref{bae})
as
\begin{eqnarray}
r_2(\l_k)\prod_{\scriptstyle j=1\atop
\scriptstyle j\neq k}^{N} {h(\lambda_k,\lambda_j)\over
h(\lambda_j,\lambda_k)} &=& {(-1)^{{N}-1}\over r_1(\l_k)}\ ,\cr
{1\over r_2(\l_k)}\prod_{\scriptstyle j=1\atop
\scriptstyle j\neq k}^{N} {h(\lambda_j,\lambda_k)\over
h(\lambda_k,\lambda_j)} &=& (-1)^{{N}-1} r_1(\l_k)\ , k=1,\ldots
,{N}
   \label{bae3}
\end{eqnarray}

we find that the additional factors take the form
\begin{equation}
   \exp\left(\kappa(\l_k)-\kappa(\l_j)\right){r_2(\l_k)\over
    r_2(\l_j)}= {r_1(\l_j)\over r_1(\l_k)}\ .
\end{equation}
To get the diagonal matrix elements we have to investigate the limit
$\l_j\rightarrow\l_k$ of (\ref{genfu8}) in detail. In the limit
$\l_j\rightarrow\l_k$ the sum of the first two terms in $G_{jk}$ and
the expression in brackets are both of the form ``${0\over 0}$''. By
using l'Hospital's rule we find
\begin{eqnarray}
&&\lim_{\l_j\rightarrow\l_k}\left(t(\l_j,\l_k) + t(\l_k,\l_j)
{r_2(\l_k)\over r_2(\l_j)}e^{\varphi(\l_k) -   \varphi(\l_j)}
e^{\kappa(\l_k) -\kappa(\l_j)}\right) =-2\cosh(2i\eta) \cr
&&+ \sinh(2i\eta){\partial \varphi(\l)\over\partial\l}
\bigg|_{\l=\l_j}+\sinh(2i\eta) {\partial\over\partial\l_j}\left[
\ln(r_2(\l_j))+\sum_n\ln({h(\l_j,\l_n)\over h(\l_n,\l_j)})\right] .
\end{eqnarray}
Putting now everything together we obtain the following representation
of the FSFP as a ratio of two determinants
\begin{eqnarray}
P(m) &=& \frac{\nddv \det{{\cal G}}\dv}{\det {\cal N}^\prime}\ ,\cr
{{\cal G}}_{jk}&=& t(\l_j,\l_k) + t(\l_k,\l_j) {r_1(\l_j)\over
r_1(\l_k)}\exp\left(\varphi(\l_k) - \varphi(\l_j)\right) \cr
&&+ \delta_{jk}\sin(2\eta)\left(L D(\l_j) +\sum_n K(\l_j,\l_n)\right)\ ,
\label{genfu6}
\end{eqnarray}
where $\varphi$ and ${\cal N^\prime}$ are defined in
(\ref{ndf2}) and (\ref{norm3}) respectively and where $D$ and $K$ are
defined in (\ref{kernel}).
Eqn (\ref{genfu6}) is our final result for the FSFP on a {\sl finite}
chain of length $L$. As always in the Bethe Ansatz things simplify
essentially in the thermodynamic limit. Let us first investigate
the thermodynamic limit for the norm $\sigma_{N}$ (\ref{norm3}). We
write ${\cal N}^\prime$ as the product of two matrices:
\begin{equation}
   {{\cal N}^\prime}_{jk} =  \sin(2\eta)
         \sum_m I_{jm} J_{mk}, \quad
   I_{jm} = \delta_{jm} - \frac{K_{jm}}{\theta_m},\quad
   J_{jm} = \delta_{jm}\theta_m,
\end{equation}
where $\theta_m = LD(\l_m)+\sum_n K(\l_m,\l_n)$. The determinant of ${\cal
N}^\prime$ is the product of the determinants of $I$ and $J$.  Next we use
that the set of roots $\{\l_j\}$ describes the ground state and the roots
thus obey the equations
\begin{equation}
   2\pi L\rho(\l_j) -\sum_{k=1}^{N} K(\l_j,\l_k) = L D(\l_j)\
,\qquad   j=1\ldots {N}\ ,
\end{equation}
which is the discrete version of (\ref{gsie}). Here $\rho(\l_j) =
\frac{1}{L(\l_{j+1}-\l_j)}$, which becomes $\rho(\l)$ defined by
(\ref{gsie}) in the thermodynamic limit. We thus can rewrite
$\theta_m=2\pi L\rho(\l_m)$, which leads to
\begin{equation}
 \det J = \prod_{j=1}^{N} 2\pi L\rho(\l_j)\ .
\label{J}
\end{equation}
In the thermodynamic limit the matrix $I$ turns into an integral operator
${\widehat I}=id -{1\over 2\pi}{\widehat K}$
\begin{equation}
   {\widehat I}*f\bigg|_\l = f(\l)-\frac{1}{2\pi}\int_{-\Lambda}^\Lambda
   d\mu K(\l,\m) f(\m)\ ,
\end{equation}
where $K$ is the kernel of $\widehat K$ defined by (\ref{kernel}).
The matrix $G_{jk}$ in (\ref{genfu6}) is treated in a very similar
way. We rewrite it as a product
\begin{equation}
   G_{jk} = \sin(2\eta) \sum_m W_{jm} J_{mk}\ ,
\end{equation}
where $J_{jm}=\delta_{jm}2\pi L\rho(\l_m)$ is the same as above, and
\begin{equation}
W_{jk}= \delta_{jk} + {1\over\sin(2\eta)\theta_k} \bigg\{
t(\l_j,\l_k) + t(\l_k,\l_j) {r_1(\l_j)\over
r_1(\l_k)}\exp\left(\varphi(\l_k) - \varphi(\l_j)\right) \bigg\}\ .
\end{equation}
In the thermodynamic limit the matrix $W_{jk}$ turns into an integral
operator ${\widehat W}= id + {1\over 2\pi}{\widehat V}$ with kernel
\begin{equation}
V^{(m)} (\l, \m)  =  \frac{-\sin(2\eta)}{\sinh(\l-\m)} \bigg\{ {1\over
\sinh(\l-\m + 2i\eta)} - {e^{-1} (\l) e(\m) \over\sinh(\m - \l +
2i\eta)}\bigg\}\ ,
\label{V}
\end{equation}
\begin{equation}
e (\l) = \bigg( {\sinh(\l + i\eta) \over \sinh(\l - i\eta)}
\bigg)^{m} \exp(\varphi (\l )) \ .
\end{equation}
Thus in the thermodynamic limit the FSFP $P(m)$ for the XXZ magnet
can be represented as a ratio of determinants of Fredholm integral
operators $\left(id + {1\over 2\pi}\widehat{V}\right)$ and
$\left(id\ -{1\over 2\pi}\widehat{K}\right)$ in the following way
\begin{equation}
P(m)= {\nddv\det (id\ +{1\over 2 \pi}\widehat{V})\dv\over \det (id\ -
{1 \over 2\pi} \widehat{K})} \ .
   \label{genfu9}
\end{equation}
Here $\nddv$ and $\dv$ are the vacua of the dual bosonic Fock space
defined in (\ref{ndf}).

\section{Integro-Difference Equations}

In this section we embed the numerator of the r.h.s. of (\ref{genfu9})
in a system of integro-difference equations. The denominator is
independent of the distance $m$ and thus merely amounts to an overall
normalization of $P(m)$, which is difficult to determine anyhow and
will be dropped in what follows. We start by bringing the kernel of
$\widehat{V}$ (\ref{V}) to ``standard'' form\citeup{vladb}.
This is done in several steps. We first perform the similarity
transformation ${\hat S}$ with kernel $S(\l)=(e(\l))^\frac{1}{2}$ on
${1+\frac{1}{2\pi}{\hat V}}$, which leaves the determinant unchanged.
This ``symmetrizes'' the kernel
\begin{equation}
V^{(m)} (\l, \m)  =  \frac{-\sin(2\eta)}{\sinh(\l-\m)} \bigg\{ {
{\bar e}^{-1} (\m) {\bar e}(\l) \over \sinh(\l-\m + 2i\eta)} -
{{\bar e}^{-1} (\l) {\bar e}(\m) \over\sinh(\m - \l + 2i\eta)}\bigg\}\ ,
\end{equation}
where
\begin{equation}
{\bar e}(\l) = \bigg( {\sinh(\l + i\eta) \over \sinh(\l - i\eta)}
\bigg)^{\frac{m}{2}} \exp(\frac{\varphi (\l )}{2}) \ .
\end{equation}
Next we change variables to $\beta=\exp(2\l)$, $\alpha=\exp(2\m)$,
$w=\exp(2i\eta)$. Elementary calculations yield
\begin{eqnarray}
&&{1\over 2\alpha}V(\frac{\ln(\beta)}{2},\frac{\ln(\alpha)}{2}) =
\frac{-2\sin(2\eta)\beta}{\beta-\alpha}\bigg\{\left(\frac{\beta
w-1}{\beta -w}\right)^{m\over 2} \left(\frac{\alpha w-1}{\alpha
-w}\right)^{-{m\over 2}}\times \cr
&&\times \frac{1}{\beta w-{\alpha\over w}}
\exp\left(\frac{\varphi(\frac{\ln(\beta)}{2})}{2} -
\frac{\varphi(\frac{\ln(\alpha)}{2})}{2} \right) -
\alpha\leftrightarrow\beta\bigg\}.
\label{valpha}
\end{eqnarray}
Next we use the identity
\begin{equation}
\int_0^\infty ds\ \exp\left(-is(\beta w-{\alpha\over w})\right) =
\frac{-i}{\beta w-{\alpha\over w}}\ ,
\end{equation}
(which holds as $\alpha>0$, $\beta>0$ and $\sin(2\eta)<0$) in order to
eliminate the unwanted denominators in (\ref{valpha}), and finally
perform yet another change of variables to $z(\alpha)=\frac{\alpha
w-1}{\alpha -w}$. This change of variables maps the real axis on the
contour $C: \alpha\to z=\exp{i\alpha}$ where $-\psi < \alpha
<2\pi+\psi\le\eta$ ($\psi<0$ by definition).  The endpoints $\xi =
e^{i\psi}$ and ${\xib} = e^{-i\psi}$ (we integrate from $\xib$ to
$\xi$) of the contour are related to the magnetic field $h$ {\em and}
the anisotropy $\Delta$. After elementary computations we obtain
\begin{equation}
{d\alpha\over 2\alpha}V(\frac{\ln(\beta)}{2},\frac{\ln(\alpha)}{2}) =
\frac{dz_2}{i}\left(\frac{wz_1-1}{wz_2-1}\frac{z_1-w}{z_2-w}\right)
\int_0^\infty ds\ {e^{(m)}_+(z_1|s) e^{(m)}_-(z_2|s) -
{z_1\leftrightarrow z_2}\over z_1-z_2}\ ,
\label{bla}
\end{equation}
where the functions $e_\pm^{(m)}$ are given by
\begin{eqnarray}
   e_+^{(m)}(z|s) &=&
     \left( 2\sin(2\eta)\ {wz-1\over z-w}\ z^{-m}e^{-\phi(z)}\right)^{1\over2}
     \exp\left( i{s\over w} {wz-1\over z-w}\right)\nonumber \\
   \label{def:e} \\
   e_-^{(m)}(z|s) &=&
     \left( 2\sin(2\eta)\ {wz-1\over z-w}\ z^{m}e^{\phi(z)}\right)^{1\over2}
     \exp\left( -isw{wz-1\over z-w}\right)\nonumber .
\label{e}
\end{eqnarray}
Note that the ``momenta'' $p(z)$ and ``coordiantes'' $q(z)$ of the
dual field $\phi(z)$ now obey the commutation relations
\begin{equation}
[q(z_2),p(z_1)]=\ln\left[\frac{1}{4\sin^2(2\eta)}\left(
\frac{wz_2-1}{wz_1-1}\frac{z_1-w}{z_2-w}+\frac{wz_1-1}{wz_2-1}
\frac{z_2-w}{z_1-w} -w^2-w^{-2}\right)\right] .
\end{equation}
Eliminating the factor in brackets before the integral in (\ref{bla})
by a similarity transformation we obtain
\begin{eqnarray}
\det(1+{1\over 2\pi}{\hat V}) &=& \det(1+{\hat V}^{(m)})\cr
V^{(m)}(z_1|z_2)&=& {i\over 2\pi}\int_0^\infty ds\
             {e^{(m)}_+(z_1|s) e^{(m)}_-(z_2|s)
            - e^{(m)}_-(z_1|s) e^{(m)}_+(z_2|s)
             \over z_1-z_2}\ .
   \label{vkernelz}
\end{eqnarray}
The integral operator $\widehat{V}^{(m)}$ now acts on a function
$f(z)$ as
\begin{equation}
   \left( \widehat{V}^{(m)} f \right) (z_1)
                = \int_C dz_2 V^{(m)}(z_1|z_2) f(z_2) ,
\end{equation}
where the integration is to be performed along the contour
$C$. We note that $V$ is symmetric and nonsingular at $z_1=z_2$.
Eqn (\ref{vkernelz}) is the desired standard form for the kernel, and
was obtained for the price of introducing the auxiliary s-integration.

The resolvent $\widehat{R}^{(m)}$ of $\widehat{V}^{(m)}$ is defined by
$(1+\widehat{V}^{(m)})(1-\widehat{R}^{(m)})=1$
and its kernel $R^{(m)}(z_1|z_2)$ can be written in a form
similar to Eq.~(\ref{vkernelz}), namely \citeup{vladb}
\begin{equation}
   R^{(m)}(z_1|z_2) = -{i\over{2\pi}}\int_0^\infty ds\ {{f^{(m)}_+(z_1|s)
f^{(m)}_-(z_2|s) - f^{(m)}_-(z_1|s) f^{(m)}_+(z_2|s)} \over {z_1-z_2}} \; .
  \label{rkernelz}
\end{equation}
Here the functions $f_\pm^{(m)}$ are solutions of the linear integral
equations
\begin{equation}
   f_\pm(z_1|s) + \int_C dz_2 V(z_1|z_2) f_\pm(z_2|s)
         = e_\pm(z_1|s) \; .
   \label{def:f}
\end{equation}
In terms of these functions we introduce the integral operators
$B_{ab}^{(m)}$, $a,b=\pm$ acting as $(B^{(m)}_{ab} f^{(m)})(s) =
\int_0^\infty dt\ B^{(m)}_{ab}(s,t) f(t)$ with the kernel
\begin{equation}
  B^{(m)}_{ab}(s,t) = {i\over2\pi} \int_C {dz\over z} f^{(m)}_a(z|s)
e_b^{(m)}(z|t)\, , \qquad a,b = \pm \, .
  \label{def:B}
\end{equation}
The {\em transpose} $B^T$ acts like $\left(B^T f\right)(s) = \int_0^\infty
dt\ B(t,s) f(t)$. We are now able to formulate the embedding of $P(m)$
into a set of integrable integro-difference equations:
\begin{itemize}
\item[(i)]
The lattice logarithmic derivative of $\det\left( 1+\widehat{V}^{(m)}
\right)$ is given in terms of the integral operator $B^{(m)}_{ab}$ as
\begin{equation}
   { {\det\left( 1+ \widehat{V}^{(m+1)} \right)} \over
     {\det\left( 1+ \widehat{V}^{(m)} \right)} }
   = \det\left( 1 + B_{-+}^{(m)} \right)
   \label{vvd1}
\end{equation}
\item[(ii)]
The logarithmic derivative of $\det\left(1+\widehat{V}^{(m)}\right)$ with
respect to the boundaries of the contour $C$ is expressed in terms of the
functions $F_\pm^{(m)}(s)=f_\pm^{(m)}(\xi|s)$ and
$G_\pm^{(m)}(s)=f_\pm^{(m)}({\bar\xi}|s)$ as follows
\begin{eqnarray}
   -i \partial_\psi \ln \det\left( 1+\widehat{V} \right) &=& {1\over 2\pi}
     \int_0^\infty ds\ \Bigl\{F_+(s) \partial_\psi F_-(s)
                 - F_-(s) \partial_\psi F_+(s)
       \nonumber \\
  &&\qquad \qquad
                 + G_-(s) \partial_\psi G_+(s)
                 - G_+(s) \partial_\psi G_-(s)
            \Bigr\} \nonumber \\
  &&+{1\over 4\pi^2} {\xi+\xib \over \xi-\xib} \;
        \left( \int_0^\infty ds\ F_+(s) G_-(s)
                    - F_-(s) G_+(s) \right)^2 \; .
  \label{eq:vvdpsi}
\end{eqnarray}
\item[(iii)]
The following set of completely integrable integro-difference equations for
the unknowns $F,G,B$ in (i) and (ii) holds
\begin{eqnarray}
   {1 \over \sqrt{\xi}} F_+^{(m+1)} &=&\! {1\over \xi}
        \left\{ F_+^{(m)}
          - \left( 1- B_{+-}^{(m+1)} \right)
            \left( B_{++}^{(m)} \right)^T F_-^{(m)} \right\}
   \nonumber \\
   {1 \over \sqrt{\xi}} F_-^{(m+1)} &=&\!
      {1\over \xi} \left\{
        \left( \xi + B_{--}^{(m+1)} \left(B_{++}^{(m)}\right)^T
        \right) F_-^{(m)}
        - B_{--}^{(m+1)} \left( 1+ \left(B_{+-}^{(m)}\right)^T
                         \right) F_+^{(m)} \right\}
   \nonumber \\
   \label{result:f} \\
   {1 \over \sqrt{\xib}} G_+^{(m+1)} &=&\! {1\over \xib}
        \left\{ G_+^{(m)}
          - \left( 1- B_{+-}^{(m+1)} \right)
            \left( B_{++}^{(m)} \right)^T G_-^{(m)} \right\}
   \nonumber \\
   {1 \over \sqrt{\xib}} G_-^{(m+1)} &=&\!
      {1\over \xib} \left\{
        \left( \xib + B_{--}^{(m+1)} \left(B_{++}^{(m)}\right)^T
        \right) G_-^{(m)}
        - B_{--}^{(m+1)} \left( 1+ \left(B_{+-}^{(m)}\right)^T
                         \right) G_+^{(m)} \right\}
   \nonumber
\end{eqnarray}
and
\begin{eqnarray}
   -i {\partial \over \partial \psi} B_{ab} (s,t) &=&
    {i\over 2\pi} \Biggl\{ 
      F_a(s) \left[ F_b(t)
                      + \left( F_+ B_{-b} - F_- B_{+b} \right) (t)
                 \right] \nonumber \\
   && \qquad +
      G_a(s) \left[ G_b(t)
                      + \left( G_+ B_{-b} - G_- B_{+b} \right) (t)
                 \right] \Biggr\}
   \label{B:dpsi}
\end{eqnarray}
where $a,b=\pm$.
\end{itemize}
The proof of (i)-(iii) is analogous to the one for the XXX-case
\citeup{fik} so that we only sketch the main steps.
(i) is a direct consequence of the {\em shift-equation}
\begin{equation}
   z_1^{-{1\over2}}\; V^{(m+1)}(z_1|z_2)\; z_2^{1\over2}
      = V^{(m)}(z_1|z_2)
      + {i\over2\pi}\int_0^\infty ds\ {e_+^{(m)}(z_1|s)
e_-^{(m)}(z_2|s) \over z_1}\ . \label{shift:v}
\end{equation}
which follows directly from (\ref{V}) and (\ref{e}), whereas
(ii) and (iii) are consequences of the Lax-representation
\begin{eqnarray}
{1\over\sqrt{z}} f_-^{(m+1)}(z|s) &=& f_-^{(m)}(z|s) - {1\over z}
B_{--}^{(m+1)} \left( \left[1+ \left(B_{-+}^{(m)}\right)^T\right]
f_+^{(m)} - \left(B_{++}^{(m)}\right)^T f_-^{(m)} \right)(z|s)
   \nonumber \\
{1\over\sqrt{z}} f_+^{(m+1)}(z|s) &=& {1\over z} f_+^{(m)}(z|s)
  -\left(\left[ 1-B_{+-}^{(m+1)}\right] \left(B_{++}^{(m)}\right)^T
f_-^{(m)} \right)(z|s)\ ,
   \label{shift:fright}
\end{eqnarray}
and
\begin{eqnarray}
   {\partial \over \partial \psi} f_\pm(z|s) &=&
     {1\over 2\pi}\, \left\{ {\xi\over z-\xi} f_\pm(\xi|s)
              \int_0^\infty dt\ \left(f_+(\xi|t) f_-(z|t)
                       - f_-(\xi|t)f_+(z|t)\right)
                    \right. \nonumber \\
  && \quad \left.
      + {\xib\over z-\xib} f_\pm(\xib|s)
              \int_0^\infty dt\ \left(f_+(\xib|t) f_-(z|t)
                       - f_-(\xib|t)f_+(z|t)\right) \right\}.
  \label{f:dpsi2}
\end{eqnarray}

This concludes our derivation of the embedding of $P(m)$ in a system
of integrable (due to the Lax-representation
(\ref{shift:fright})-(\ref{f:dpsi2})) integro-difference equations. We
note that there is a integral-operator-valued Riemann-Hilbert problem
associated with these IDE\cite{IDE}, and that the asymptotics of
$P(m)$ for large $m$ can be extracted from the solution of the
Riemann-Hilbert problem. So far this has been done only for some
limiting cases\cite{IDE}, but we hope to report the full solution in
the future.

\section{Acknowledgements}
V.E.K. thanks the organizers of the conference for support.

\end{document}